\documentclass[12pt]{iopart}
\usepackage[latin1]{inputenc}
\usepackage{graphicx}

\newenvironment{Glossarylist}{%
  \begin{list}{}{%
      \setlength{\labelsep}{1em}%
      }%
}%
{\end{list}}

\newcommand{\chem}[1]{\ensuremath{\mathrm{#1}}}
\newcommand{\un}[1]{\ensuremath{\unskip\,\mathrm{#1}}}

\begin{document}

\title[Effect of channel block  on the spiking activity]{Effect of
channel block on
the spiking activity of excitable membranes in
a stochastic Hodgkin-Huxley model}
\author{G. Schmid,
I. Goychuk and P. H\"anggi\footnote[7]{Corresponding author:
hanggi@physik.uni-augsburg.de}}

\address{Institut f\"ur Physik, Universit\"at Augsburg,
D-86135 Augsburg, Germany}

\begin{abstract}
The influence of intrinsic channel noise on the spontaneous
spiking activity of poisoned excitable membrane patches is studied
by use of a stochastic generalization of the Hodgkin-Huxley model.
Internal noise stemming from the stochastic dynamics of individual
ion channels is known to affect the collective properties of the
whole ion channel cluster. For example, there exists an optimal
size of the membrane patch for which the internal noise alone
causes a regular spontaneous generation of action potentials.
In addition to varying the size of ion channel clusters, living
organisms may adapt the densities of ion channels in order to
optimally regulate the spontaneous spiking activity.
The influence of channel block on the excitability of a membrane patch
of certain size is twofold: First, a variation of ion channel
densities primarily yields a change of the conductance level.
Second, a down-regulation of working ion channels always increases the
channel noise. While the former effect dominates in the case of
sodium channel block resulting in a reduced spiking activity,
the latter enhances the generation of spontaneous action potentials
in the case of a tailored potassium channel blocking.
Moreover, by blocking some portion of either potassium or sodium ion channels, it
is possible to either increase or to decrease the regularity of the
spike train.
\end{abstract}


\submitto{Physical Biology}


\section{Introduction}
\label{sec:introduction}

Following the study of Hodgkin and Huxley \cite{Hodgkin1952}, most
of the models of axons have treated the generation and propagation
of action potentials using deterministic  differential equations.
Since the work of Lecar and Nossal \cite{Lecar1971} it became clear,
however,
that not only the synaptic noise but also the randomness of the ion channel
gating itself may  cause
threshold fluctuations in neurons \cite{White2000}.
Therefore, channel noise which
stems from the stochastic nature of the ion channel dynamics
must be taken into account \cite{White2000}. It impacts such
features as the threshold to spiking and the spiking rate itself
\cite{Skaugen,Clay1983,Strassberg1993,DeFelice1993,Fox1994,Chow1996,Schneidman},
the anomalous noise-assisted enhancement of transduction of external signals, i.e.
the phenomenon of
Stochastic Resonance \cite{Schmid2001,Jung2001,Gammaitoni1998,Hanggi2002} and the efficiency for
synchronization \cite{schmid2003}, to name but a few such phenomena.
The origin of the channel noise \cite{White2000} is
basically due to fluctuations of the mean number of open ion
channels around the corresponding mean values.
Therefore, the strength of the channel
noise is mainly determined by the number of ion channels
participating in the generation of action potentials.
Interestingly, there is an optimal patch sizes for which the spike
production becomes more regular \cite{Schmid2001,Jung2001}. The
objective of this work is to investigate how the regularity of the
spiking can possibly be controlled for a given membrane patch size.
Toxins like tetraethylammonium
(TEA) and tetrodotoxin (TTX) allow to reduce
the number of working potassium or sodium ion channels, respectively,
for an extended period of time \cite{Hille2001}.
Moreover, the densities of ion channels can be adapted by
the living cell also dynamically  \cite{LeMasson,Stemmler}
over an extended time span which is needed, e.g., to express
the required ion channel proteins in the membrane \cite{Stemmler}.
The effect of blocking the ion
channels entails several different tendencies at the same time. For sodium ion channels,
the reduction of the density
of channels results in an increase of the activation
barrier towards excitation from the resting state and, therefore, in the
reduction of neuronal activity. On the other hand, however,
the corresponding channel noise component
will also be increased due to the reduction of the absolute
number of ion channels
in the membrane patch.
As a consequence, the increased channel noise will
help to overcome the activation barrier
and to initiate spontaneous spikes.
A reduction of the density of potassium channels will on the contrary
generally result in a lowering of
the activation barrier (an increase of the excitability -- see, e.g.,
in \cite{Goldman}) and, simultaneously,
in an increase of the recovery time which should favor longer interspike
time intervals. Moreover, the reduction of the total number of potassium
ion channels will also increase the corresponding channel noise component
which is expected to lead to an increased variability of the refractory period.
Which of these various concurrent effects will dominate
is not clear {\it a priori}. This depends both on the ion channel
densities and on the size of the studied membrane patch. It is the main
objective of this paper to study this highly nontrivial, subtle issue in a
stochastic model which extends the Hodgkin-Huxley model of neuronal
excitability.

\section{A stochastic Hodgkin-Huxley model}
\label{sec:hhmodel}

According to the
Hodgkin-Huxley model the dynamics of the membrane potential $V$ , measured throughout
this work in mV is
given by:
\begin{equation}
   \label{eq:voltage-equation}
   \fl C \frac{\mathrm{d}}{\mathrm{d}t} V +G_{\chem{K}}(n)\ (V-E_{\chem{K}})
   +G_{\chem{Na}}(m,h)\ ( V - E_{\chem{Na}})
    +G_{\chem{L}}
   \ (V - E_{L}) = 0\, .
\end{equation}
In Eq.~(\ref{eq:voltage-equation}), $C=1\un{\mu F/cm^2}$ is
the capacity of the cell membrane. Furthermore,
$E_{\chem{Na}}=50\un{mV}$, $E_{\chem{K}}=-77\un{mV}$
and $E_{\mathrm{L}}=-54.4\un{mV}$ are the
reversal potentials for the potassium, sodium and leakage currents,
correspondingly.
While the leakage conductance is assumed to be constant,
$G_{\mathrm{L}} =0.3\un{mS/cm^2}$, the potassium and sodium
conductances read:
\begin{equation}
  \label{eq:conductances-hodgkinhuxley}
  G_{\chem{K}}(n)=g_{\chem{K}}^{\mathrm{max}}\ x_{\chem{K}}\ n^{4} , \quad
  G_{\chem{Na}}(m,h)=g_{\chem{Na}}^{\mathrm{max}}\ x_{\chem{Na}}\ m^{3} h\, ,
\end{equation}
where {$g_{\chem K}^{\mathrm{max}}=36\un{mS/cm^2}$} and
{$g_{\chem{Na}}^{\mathrm{max}}=120\un{mS/cm^2}$} denote the
maximal conductances (when  all the channels are open). In
Eq.~(\ref{eq:conductances-hodgkinhuxley}) we introduce the factors
$x_{\chem{K}}$ and $x_{\chem{Na}}$ which are the
fractions of working, i.e., non-blocked ion channels, to the
overall number of potassium, $N_{\chem{K}}$, or sodium,
$N_{\chem{Na}}$, ion channels, correspondingly. These factors are
confined to the unit interval. Experimentally, they can be
controlled by adding cell toxins like  tetraethylammonium (TEA)
and/or tetrodotoxin (TTX) which completely block and disable
potassium or sodium ion channels, respectively \cite{Hille2001}.

While the gating variables $n,\ m$ and $h$ describes the mean
ratios of the open gates of the working channels, the factors
$n^{4}$ and $m^{3}\ h$ are the mean portions of the open ion
channels within a membrane patch. This follows from the fact that
the gating dynamics of each ion channel is assumed to be governed
by four independent gates each of which can switch between an
open and a closed conformation. The voltage-dependent opening and
closing rates $\alpha_x(V)$ and $\beta_x(V)\; (x=m,h,n)$, read
\cite{Schmid2001,Jung2001}:
\numparts
\label{eq:rates}
\begin{eqnarray}
  \label{eq:rates1}
  \alpha_{m}(V) &= \frac{ 0.1\ ( V + 40 )}{1-\exp [ - ( V + 40 ) /
    10] },\\
  \beta_{m}(V) &= 4 \ \exp [ - ( V + 65 ) / 18 ]\, ,  \\
  \alpha_{h}( V ) &=  0.07 \ \exp [ - ( V + 65 ) / 20 ], \\
  \beta_{h}( V ) &= \{ 1 + \exp [ - ( V + 35 ) / 10 ] \}^{-1}\, , \\
  \alpha_{n}( V ) &= \frac{ 0.01 \ ( V + 55 )}{ 1 - \exp [ -( V +
    55 )/10 ]},\\
  \label{eq:rates2}
  \beta_{n}( V ) &= 0.125 \ \exp [ - ( V + 65 ) / 80 ]\, .
\end{eqnarray}
\endnumparts
Fox and Lu \cite{Fox1994} have extended the Hodgkin-Huxley model
by taking into account the fluctuations of the numbers of
open ion channels around the corresponding mean values. Within
a corresponding stochastic description, the gating variables
become stochastic quantities
obeying the following Langevin equations:
\begin{equation}
  \label{eq:stochasticgates}
  \frac{\mathrm{d}}{\mathrm{dt}} x =
  \alpha_{x}(V)\ (1-x)-\beta_{x}(V)\ x + \xi_x(t), \quad x=m,h,n\, ,
\end{equation}
with independent Gaussian white noise sources $\xi_x(t)$ of vanishing mean.
For an excitable
membrane patch with $N_{\chem{Na}}$ sodium and $N_{\chem{K}}$
potassium ion channels the noise correlations assume the following form:
\numparts
\begin{eqnarray}
  \label{eq:correlator-a}
  \langle \xi_{m}(t) \xi_{m}(t') \rangle &= \frac{2}{N_{\chem{Na}}\
    x_{\chem{Na}}}\ \frac{ \alpha_{m}(V)
    \beta_{m}(V)}{[\alpha_{m}(V) +\beta_{m}(V)]}\ \delta(t -t')\, ,  \\
  \label{eq:correlator-b}
  \langle \xi_{h}(t) \xi_{h}(t') \rangle &=  \frac{2}{N_{\chem{Na}}\
    x_{\chem{Na}}}\ \frac{ \alpha_{h}(V)
    \beta_{h}(V)}{[\alpha_{h}(V) +\beta_{h}(V)]}\ \delta(t -t')\, , \\
  \label{eq:correlator-c}
  \langle \xi_{n}(t) \xi_{n}(t') \rangle &=  \frac{2}{N_{\chem{K}}\
    x_{\chem{K}}}\ \frac{ \alpha_{n}(V)
    \beta_{n}(V)} {[\alpha_{n}(V) +\beta_{n}(V)]}\ \delta(t -t')  \, .
\end{eqnarray}
\endnumparts
The overall numbers of involved potassium and sodium ion channels
are re-scaled by $x_{\chem{Na}}$ and $x_{\chem{K}}$,
respectively, in order to disregard the blocked channels which do
not contribute to the channel noise.
With the assumption of
homogeneous ion channel densities, {$\rho_{\chem{Na}} = 60\un{\mu
m^{-2}}$ and {$\rho_{\chem K} = 18\un{\mu m ^{-2}}$, the ion
channel numbers are given by: $N_{\chem{Na}}= \rho_{\chem{Na}} S,
\quad N_{\chem{K}}= \rho_{\chem{K}} S$, with $S$ being the size of
the membrane patch. The number of working ion channels, i.e. the
size of the excitable membrane patch $S$, respectively, determines
the strength of the fluctuations and thus the channel noise level.
With decreasing patch size, i.e.
decreasing number of ion channels, the noise level caused by
fluctuations of the number of open ion channels increases, cf.
Eq.~(\ref{eq:correlator-a})-(\ref{eq:correlator-c}).
It is worth noting that the It\^{o}-Stratonovich dilemma of the interpretation
of the studied system of stochastic differential equations  does not
appear since each of the noise sources $\xi_x(t)$ does not depend {\it explicitly}
on the state of the corresponding variable $x$ \cite{HanggiThomas,hanggi1980}.

\section{Poisoning in the deterministic Hodgkin-Huxley model}
\label{sec:detmodel}

Before discussing the impact of channel noise on the spontaneous
spiking activity of poisoned membrane patches we first consider
the role of channel toxins on the excitability and resting potential
of the original Hodgkin-Huxley model which neglects the fluctuations
of the number of open ion channels.
Then, the equations for the gating dynamics \cite{Hodgkin1952} reads:
\begin{equation}
  \label{eq:detgating}
  \frac{\mathrm{d}}{\mathrm{dt}} x = \alpha_{x}(V)\ (1-x)-\beta_{x}(V)\ x ,
  \quad x=m,h,n\, ,
\end{equation}
with the opening and closing rates given by
Eq.~(\ref{eq:rates1})-(\ref{eq:rates2}).
Hence Eqs.~(\ref{eq:voltage-equation}),
(\ref{eq:conductances-hodgkinhuxley}) and (\ref{eq:detgating})
form a deterministic Hodgkin-Huxley model which takes poisoning into
account. Remarkably, this set of equations corresponds to the limit of
infinitely large numbers of ion channels within the stochastic generalization
which was introduced in Sect.~\ref{sec:hhmodel}.

\begin{figure}[t]
  \centering
  \includegraphics{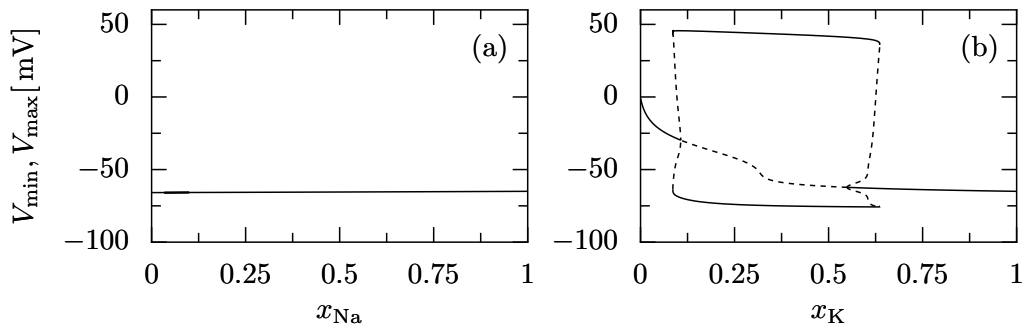}
  \caption{The bifurcation diagrams for the undriven original Hodgkin-Huxley
  model are plotted versus the ratio of intact sodium channels $x_{\chem{Na}}$,
  Fig.~(a), and the ratio of intact potassium channels $x_{\chem{K}}$, Fig.~(b),
  respectively, as bifurcation parameters. The solid line in Fig.~(a)
  denotes the stable fixed point solution. For potassium channel
  blocking, Fig.~(b), the stable fixed point solution (single solid
  line) becomes unstable with decreasing $x_{K}$, and a stable spiking
  and oscillatory solution arises (the two solid lines refer to the
  maximal and minimal membrane potential). The dashed lines
  correspond to instable solutions.}
  \label{fig:bifurcation}
\end{figure}

Fig.~\ref{fig:bifurcation} depicts the bifurcation scenario which is
derived from the deterministic model for the case of poisoning the
membrane cluster. While poisoning the sodium channels causes
only a small, practically negligible variation of the resting voltage,
see Fig.~\ref{fig:bifurcation}(a). Obviously, the poisoning of sodium channels
 can -- upon neglecting the role of channel noise -- not induce spiking events.
In clear contrast, a reduction of the number of working
potassium channels changes dramatically the qualitative behavior
of the spiking activity, see Fig.~\ref{fig:bifurcation}(b). Upon
decreasing the number of intact potassium channels a sub-critical
Hopf-bifurcation takes place -- a stable spiking and oscillatory
solution arises and the stable, non-spiking solution
becomes unstable.
Moreover, there is
region between $0.549 \leq x_{\chem{K}} \leq 0.636$, where a stable
spiking and a stable fixed point solution coexists.
With further reduction of the ratio $x_{\chem{K}}$ the oscillatory spiking
solution
loses stability, a sub-critical Hopf-bifurcation takes place and a
stable fixed point solution arises again. Once more, a region of
bistability within $0.0859 \leq x_{\chem{K}} \leq 0.1068$ is
identified.

\section{The mean interspike interval}
\label{sec:menafiring}

The numerical integration of the stochastic generalized
Hodgkin-Huxley model, cf.
Eqs.~(\ref{eq:voltage-equation})-(\ref{eq:correlator-c}), is carried
out by the standard Euler algorithm with a step size of $1\un{\mu
s}$. The Gaussian random numbers are generated by the ``Numerical
Recipes'' routine {\it ran2} using the Box-Muller algorithm
\cite{press1992}. To
ensure the confinement of the gating variables between $0$ (all
gates are closed) and $1$ (all gates are open) we have implemented
numerically reflecting boundaries at $0$ and $1$. The
occurrences of action potentials are determined by upward
crossings of the membrane potential $V$ of a certain detection
threshold. Due to the very steep increase of membrane potential at
firing the actual choice of the detection threshold does not
affect the results. In our simulations the spontaneous spikes are
found by upward crossings at zero threshold voltage. The
occurrences of action spikes $t_{i}$, $i=1,...,N$ form a point
process.

While in the original, deterministic Hodgkin-Huxley model the
action potentials occur only for a certain external current
stimulus, the intrinsic channel noise initiates spontaneous spikes
\cite{White2000,Skaugen,Clay1983,Strassberg1993,DeFelice1993,Chow1996}.
The mean interspike interval, i.e.,
\begin{equation}
  \label{eq:meaninterspike}
  \langle T \rangle = \frac{1}{N} \sum_{i=1}^{N} (t_{i}-t_{i-1})\, ,
\end{equation}
with $t_{0}=0$, becomes a function of the patch size $S$. The inverse
mean interspike interval defines the spiking rate. With increasing
noise level or decreasing patch sizes $S$, respectively, the spike
production increases and thus the mean interspike interval
$\langle T\rangle$ decreases and can approach the refractory time
\cite{Schmid2001,Jung2001}.

\begin{figure}[t]
  \centering
  \includegraphics{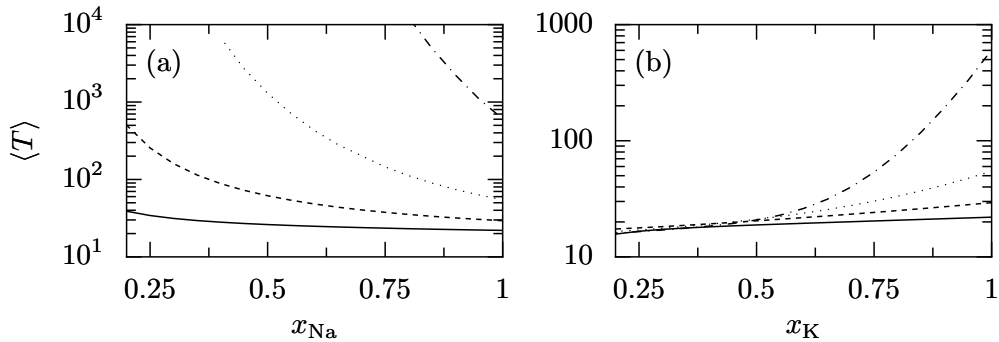}
  \caption{Mean interspike interval for poisoning: The dependence of the mean time between two
  subsequent, spontaneous action potentials versus (a) the ratio of intact sodium
  channels and (b) versus the fraction of active
   potassium  ion channels is shown for four different patch
  sizes: $S=1 \un{\mu m^{2}}$ (solid line), $S=4 \un{\mu m^{2}}$
    (dashed line), $S=16 \un{\mu m^{2}}$ (dotted line), and $S=64
  \un{\mu m^{2}}$ (dashed-dotted line).}
  \label{fig:mean}
\end{figure}

In Fig.~\ref{fig:mean} we depict the mean interspike interval
$\langle T\rangle$ against the fractions $x_{\chem{Na}}$ and
$x_{\chem{K}}$ of working sodium channels or potassium ion
channels, respectively, for different patch sizes $S$. A reduction
of the number of working sodium ion channels, i.e. smaller
$x_{\chem{Na}}$-values increases the strength of
channel noise which is caused by the stochastic behavior of the
sodium ion channels. Because the channel noise,
which is induced by the sodium channels,
is seemingly mainly responsible for the initiation of action
potentials from the rest potential \cite{Chow1996} one might expect that a
reduction of the number of sodium ion channels could then lead to
more spikes. Concurrently, however, a reduction of the number
of working sodium channels causes a diminishment of the maximal sodium
conductance. Given this
competition between these two mechanisms it is
the latter effect that dominates and, consequently, causes an
increase of the mean interspike interval, cf. Fig.~\ref{fig:mean}
(a).

The reduction of potassium conductance by poisoning the potassium
channels changes dramatically the dynamics of the original, deterministic
Hodgkin-Huxley model which ignores the impact of channel noise.
With decreasing potassium conductance a stable spiking solution
emerges at $x_{\chem{K}}=0.636$. The stable silent solution coexists
with the stable spiking solution until $x_{\chem{K}}=0.549$.
Depending on the initial value the deterministic system spends the
time either in the silent state, or on the limit cycle for
$0.549<x_{\chem{K}}<0.636$.
Consequently, the noisy system undergoes
noise-induced stochastic temporal transitions between both stable manifolds.
The motion on the spiking limit cycle becomes randomized.
For $x_{\chem{K}}<0.549$,
the stable fixed point disappears and the dynamics takes place on a
stochastic limit cycle manifold. Then, the noise does not seem to
affect strongly the mean period of the stochastic cycling. For different
patch sizes this mean period $\langle T\rangle$ is about the same,
cf. Fig.~\ref{fig:mean}(b).

However, before the Hopf-bifurcation occurs, the increase of
the {\it potassium}
channel noise can
enhance the mean frequency of the stochastic spikes occurrences.
Here,
the mean interspike interval $\langle T
\rangle$ can be drastically decreased with a reduction of working potassium channels
$x_{\chem{K}}$, cf. Fig.~\ref{fig:mean}(b).
This behavior is rather different from that known before.
It is generally
appreciated \cite{White2000,Chow1996} that the fluctuational
{\it increase}
of the number of {\it open} sodium channels is the reason for
the spontaneous spike generation. The role of the potassium channel noise
is usually thought to be negligible.
However, the decrease of the potassium conductance is known to
enhance the excitability and it can even induce spikes deterministically
(see before). Therefore, the fluctuational {\it decrease} of
the number of open potassium channel can also induce spikes.
Note in this respect
that about one third of the potassium
channels is open at the resting potential. This is contrary to the sodium channels which
are almost all closed at the resting potential.
The described alternative mechanism of the
fluctuational
spike generation  can be clearly seen at work in Fig.~\ref{fig:mean}(b)
where the sodium channel noise remains unchanged for each plotted
curve.
The effect strongly depends, however,
on the size of the membrane patch. For a sufficiently large membrane
patch (top curve in Fig. 2(b)) the discussed effect is indeed very strong.
However, when
the patch size becomes small (bottom curve in Fig. 2(b)) the poisoning
of potassium channels does not cause much effect since the stochastic
dynamics is dominated in this case by the channel noise component stemming from
the stochastic dynamics of sodium channels.
These qualitative features are displayed
in Fig.~\ref{fig:mean}(b).

\section{Controlling the Coherence of Poisoned Spiking Activity}

\begin{figure}[t]
  \centering
  \includegraphics{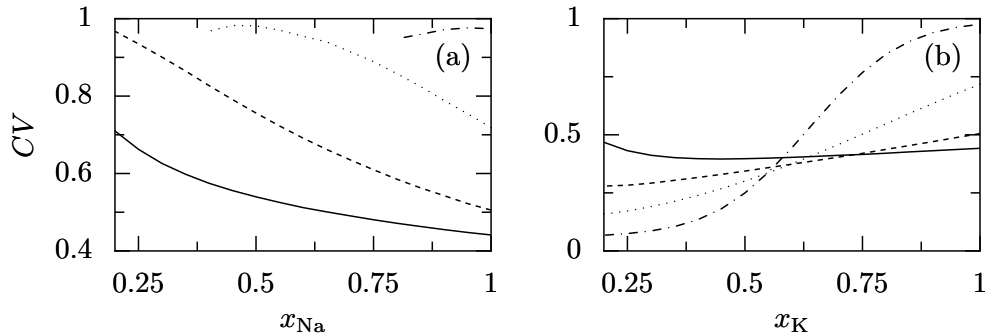}
  \caption{Same as in Fig.~\ref{fig:mean}, but for the coefficient of
    variation $CV$, cf. Eq.~(\ref{eq:cv}).
    By addition of TEA or TTX the regularity of the spiky output can
    both be increased and decreased.
    }
  \label{fig:toxic}
\end{figure}

We next address the regularity of spontaneous action potentials. A
proper measure is the coefficient of variation, $CV$, a measure of
coherence, which is given as the ratio of standard deviation to
the mean value:
\begin{equation}
  \label{eq:cv}
  CV = \frac{\sqrt{\langle T^{2}\rangle - \langle T
  \rangle^{2}}}{\langle T \rangle}\, ,
\end{equation}
where $\langle T^{2}\rangle := \frac{1}{N} \sum
(t_{i}-t_{i-1})^{2}$ is the mean-squared interspike interval. For
a fully disordered point process (the case of Poisson process) the
coefficient of variation $CV$ assumes the value $CV=1$, while for
more ordered processes it assumes smaller values and for a
deterministic signal it vanishes. In  previous
studies it has been demonstrated that $CV$ exhibits a distinct
minimum for an optimal patch size $S\approx 1\un{\mu m^{2}}$ at
which the spiking is mostly regular \cite{Schmid2001,Jung2001}.
This phenomenon has been termed intrinsic coherence resonance.

In Fig.~\ref{fig:toxic}, the coefficient of variation is plotted
{\it vs.} the fractions $x_{\chem{Na}}$ (a) and $x_{\chem{K}}$ (b).
Any addition
of TTX which blocks sodium ion channels leads to an increase of
the $CV$, cf. Fig.~\ref{fig:toxic}(a). In contrast, toxins which
disable potassium channels yield a rise of the regularity, even beyond
the level which can be reached for an optimal patch size with the unmodified
density of the ion channels, cf.
Ref.~\cite{Schmid2001}. This result is due to the fact that the lowering of
the potassium conductance level promotes a stable
oscillatory spiking solution in the deterministic model. The
channel noise, on the contrary, tends to disturb the regular spiking.
Nevertheless, the former effect dominates
resulting in the observed behavior for the $CV$.
In the case of optimal patch size, where the phenomenon of intrinsic
coherence resonance occurs, the poisoning of potassium channels
does not produce a strong effect. This is in a clear contrast with
the poisoning of sodium channels which destroys the coherence resonance
(bottom curve in Fig. 3(a)).

\section{Conclusion and outlook}
\label{sec:Conclusion}

Our study of the stochastically generalized Hodgkin-Huxley model
reveals the possibility to manipulate  the response
of a spiking membrane patch by adding toxins which selectively
block ion channels. For example, by a fine-tuned addition of
tetrodotoxin (TTX) a certain portion of sodium ion channels could be
experimentally disabled on purpose. This in turn results in a reduction of
the spontaneous action potentials and causes a more irregular
production of spikes.  On the other hand, the  addition of
tetraethylammonium (TEA) can be  used in order to block
potassium ion channels.
This causes a surprising increase of
the spiking activity (i.e. a decrease of the mean interspike
interval) and yields in turn a more regular spontaneous spiking
coherence. These characteristic features are expected to impact as well
the behavior of biological ``Stochastic Resonance'',
i.e. the phenomenon that the application of an appropriate dose of
noise can boost signal transduction
\cite{Gammaitoni1998,Hanggi2002} and, also, the phenomenon of ``Coherence
Resonance'' \cite{PikovskyKurths} in oscillatory or excitable biological entities
\cite{Schmid2001}.

We share the confident belief that our
study of the tailored control of channel noise via channel blocking
in an archetypal model of excitable biological membranes
provides some insight into the underlying principles and mechanisms and
 thus will motivate further studies of more
realistic models of real neurons where such channel noise phenomena
do play an essential and constructive role.

{\bf Acknowledgments}. This work has been supported by the Deutsche
For\-schungs\-ge\-mein\-schaft via the Son\-der\-for\-schungs\-bereich SFB-486,
project A10.

\section*{Glossary}
\begin{Glossarylist}
  \item [Action potential / spike] A rapid change of the trans-membrane
    electrical potential in the excitable cell membrane.
  \item [Bifurcation] A qualitatively change in the topology of the
    attractor-basin phase portrait under a small variation of a
    parameter within a nonlinear system.  
  \item [Channel noise] The fluctuations of the number of open ion channels.
  \item [Coherence Resonance] Noise-induced improvement of the
    regularity of the system's output. 
  \item [Excitable membrane] Cell membranes that are capable of
    rapid 
    changing their trans-membrane electrical
    potential in form of spikes.
  \item [Gates / gating] Closing and opening function of an ion
    channel. This usually refers
    to the movement of protein structural elements of the channel that
    occludes the channel pore.
  \item [Gating variable] A variable which accounts for the fraction of
    open channel gates of a certain type.
  \item [Gaussian white noise] Stochastic trajectories whose
    distributions are uncorrelated
    in time and are Gaussian-distributed.
  \item [Hopf-bifurcation] A bifurcation of a fixed point to an 
    oscillatory solution. When the oscillatory solution is unstable
    and exists at subcritical values of the control parameter, we term it
    a subcritical Hopf-bifurcation. Conversely, when the oscillatory
    solution is stable and exists at supercritical values of the
    control parameter, a supercritical Hopf-bifurcation is observable. 
  \item [Ion channel] A protein folded in the cell membrane which
    enables the transport of specific ions through the membrane.
  \item [It\^{o}-Stratonovich dilemma] Interpretation problem which arises
    in the context of Langevin equations in case of multiplicative
    Gaussian white noise. 
  \item [Langevin equation] An equation of motion that describes the temporal
    evolution of a variable which is subjected to noise acting on the
    system.   
  \item [Limit cycle] An attracting set of a nonlinear system to which
    trajectories converge. Upon the limit cycle these trajectories are
    closed and periodic.  
  \item [Manifold] 
    A topological space wherein a stable solution of a
    nonlinear system remains.
  \item [Membrane patch] A portion of the cell membrane with a certain
    size.
  \item [Membrane potential] The trans-membrane electrical potential.
  \item [Opening / closing rate] Under the two-state assumption for
    the gate dynamics (open and closed) these give the probability per
    time unit for
    transitions between the two states of a single gate.
  \item [Poisson process]  A stochastic process which is memory-less
    (Markovian) with exponentially distributed waiting times between
    two successive events.
  \item [Recovery time] The time needed for relaxation to the resting
    state after an excitation occurred. Under physiological
    conditions an initiation of an action potential within this time
    span is not possible.
  \item [Reversal potential] The electrical potential for which the
    trans-membrane flux of specific ions vanishes.
  \item [Rest potential / rest state] The
    equilibrium position of the trans-membrane potential of a
    certain membrane patch. 
  \item [Spontaneous spiking activity] The occurrence of action
    potentials which are not initiated by an externally applied stimulus.
  \item [Stochastic Resonance] A anomalous, noise-assisted enhancement
    of transduction of weak (deterministic or stochastic) signals.
\end{Glossarylist}

\end{document}